# Effects of the May 2024 Solar Storm on the Earth's Radiation Belts Observed by CALET on the International Space Station

A. Ficklin[1], A. Bruno[2,3], L. Blum[4], N. Cannady[5], T. G. Guzik[1,†], R. Kataoka[6], K. Munakata[7], Y. Akaike[8,9] and S. Torii[8]

[1]Department of Physics and Astronomy, Louisiana State University, Baton Rouge, LA, USA, [2]Heliophysics Science Division, NASA Goddard Space Flight Center, Greenbelt, MD, USA, [3]Department of Physics, Catholic University of America, Washington DC, USA, [4]Laboratory of Atmospheric and Space Physics, University of Colorado, Boulder, CO, USA, [5]Astroparticle Physics Laboratory, NASA Goddard Space Flight Center, Greenbelt, MD 20770, USA, [6]Okinawa Institute of Science and Technology, 1919-1 Tancha, Onna-son, Okinawa, 904-0495, Japan, [7]Faculty of Science, Shinshu University, 3-1-1 Asahi, Matsumoto, Nagano 390-8621, Japan, [8]Waseda Research Institute for Science and Engineering, Waseda University, 17 Kikuicho, Shinjuku, Tokyo 162-0044, Japan, [9]JEM Utilization Center, Human Spaceflight Technology Directorate, Japan Aerospace Exploration Agency, 2-1-1 Sengen, Tsukuba, Ibaraki 305-8505, Japan

**Key Points:**

- Following the extreme geomagnetic storm of early May 2024, CALET identified a new population of relativistic electrons between L~2.2-3.2

- The new electron storage ring was found to persist for more than several months, depending on energy and L

- Electron lifetimes, measured at >1.5, >3.4 and >8.2 MeV, increase with energy at L=2.2 but decrease at L=3, showing a transition in trend

**Abstract**

In May 2024, extraordinary solar activity triggered a powerful solar storm, impacting Earth and producing the extreme geomagnetic storm of May 10-11, the most intense since 2003. This had significant effects on the magnetosphere, leading to the creation of a new long-lasting component of relativistic electrons and to flux changes in the South-Atlantic Anomaly. Here we present radiation-belt observations made by the *Calorimetric Electron Telescope* (CALET) on the *International Space Station*. Specifically, we took advantage of the count rates from three layers of the CALET charge detector and imaging calorimeter. We show that the new electron storage ring extended to energies in the multi-MeV range and down to McIlwain's L=2.2, well below the nominal slot-region barrier of L=2.8, and persisted for several months, depending on energy. The evolution of the new radiation-belt configuration over time was characterized by estimating the decay rates as a function of energy and L.

**Plain Language Summary**

The radiation environment of Earth typically consists of two separate structures: the inner radiation belt centered around L=1.5 (L is the distance in Earth radii between the center of the Earth's equivalent magnetic dipole to a position at the magnetic equator), and the outer radiation belt at L$\gtrsim$4. Between the two is a slot region mostly devoid of relativistic ($\gtrsim$1 MeV) electrons. During exceptionally-strong geomagnetic storms it is possible for electrons from the outer radiation belt to be transported deep into this slot region and persist for an extended period. This occurred during the strongest geomagnetic storm in over 20 years on May 10-11, 2024. Count rates from the *Calorimetric Electron Telescope* (CALET) show that, following this event, relativistic electrons from the outer belt were pushed down to L-shell values of L$\approx$2.2-3.2 and persisted there for over 5 months depending on energy. Estimated lifetimes of these electrons provide potential insight into the loss mechanisms for relativistic electrons in this region.

**1 Introduction**

Early May 2024 featured extreme solar activity related to NOAA active regions (ARs) 13664 and 13668. These regions merged on May 3-7 near the solar central meridian, then rotated over the west limb on May 13 (Kwak et al., 2024; Li et al., 2024). This complex sunspot cluster, the largest since 1990, exhibited an extraordinary flaring activity. According to the Lockheed Martin's *Solarsoft* system, 101 M-class and 18 X-class flares were produced in rapid succession, several accompanied by Earth-directed coronal mass ejections (CMEs) based on the *Space Weather Database Of Notifications, Knowledge, Information* (*DONKI*) at the NASA *Community Coordinated Modeling Center*. In particular, simulation studies suggest that the 1130 km/s-fast CME first observed by the *Large Angle and Spectrometric Coronagraph* (LASCO) onboard the *Solar and Heliospheric Observatory* (SOHO) spacecraft on May 8 at 22:24 UT, and associated with an X1.0 flare, interacted and merged with three CMEs previously erupted from the same AR with similar directions and lower speeds. This resulted in a interplanetary CME (ICME) structure propagating to Earth (e.g., Liu et al., 2024; Hayakawa et al., 2025). Such combined ejecta, characterized by a significantly-enhanced interplanetary magnetic field (IMF) strength, impacted the dayside magnetosphere late on May 10 triggering an extreme G5 geomagnetic storm according to the NOAA's space-weather scale. Characterized by a planetary K-index (Kp) of 9 and a minimum disturbed storm-time (Dst) index of -412 nT, it can be classified as a superstorm, the largest in two decades, and the third largest of the space age after the 1989 March storm (Dst=-589 nT) and the 2003 Halloween storm (Dst=-422 nT). The 10-11th May 2024 storm ranks 9th in the 110-year history of the Kakioka Magnetic Observatory and 6th in the 66-year history of the Dst index (Kataoka et al., 2024).

This event had a significant impact on the terrestrial magnetosphere, leading to the formation of a new electron population in the outer radiation belt which was injected into the slot region, as previously reported by missions such as PROBA-V/*Energetic Particle Telescope* (EPT; Pierrard et al., 2024) and the *Relativistic Electron and Proton Telescope integrated little experiment-2* (REPTile-*2*; Li et al., 2025). The slot region, nominally extending between McIlwain's L values of ~2-3, separates the relatively-stable inner belt and the highly-variable outer belt that is often strongly affected by solar-induced disturbances (e.g., Baker et al., 2019). During geomagnetically quiet times, the slot region is typically characterized by low relativistic electron intensity resulting from competition between acceleration and loss mechanisms (Lyons & Thorne, 1973). However, it can be filled when the outer-belt electron population occasionally penetrates into the inner belt during the recovery phase of large storms (e.g., Russell and Thorne (1970); Turner et al., 2016). Generally, slot region filling occurs within hours to days of a geomagnetic disturbance, with electron intensities able to increase by orders of magnitude compared to pre-storm levels (Baker, 2004). The subsequent decay of this new low-L population takes tens to hundreds of days, depending on electron energy and L (Baker et al., 1994; Meredith et al., 2007; Ripoll et al., 2014; Reeves et al., 2015). This long-lasting depletion process can have important implications in terms of the radiation levels experienced in this region.

In this work, we present the radiation-belt observations following the 2024 May storm made by the *Calorimetric Electron Telescope* (CALET) onboard the *International Space Station* (ISS). The paper is structured as follows: CALET is introduced in section 2; in section 3 we analyze CALET measurements together with contextual interplanetary and geomagnetic data; in section 4 we investigate the lifetime of the new radiation-belt component; and finally, section 5 reports our summary and conclusions.

**2 Instrument and Data**

CALET, a Japanese-US-Italy high-energy astrophysics instrument, was launched to the ISS in August 2015 (Torii S., 2016). The primary scientific goals of the mission include investigation of the acceleration mechanisms and propagation of cosmic rays through the precise measurement of electron, proton, gamma-ray, and heavy-nuclei energy spectra extending up to 1 PeV in energy. In addition, the ISS orbit (~370-460 km altitude and ~51.6° inclination) also allows it to frequently sample the radiation environment at low-Earth orbit (LEO). This enables continuous monitoring of space-weather phenomena affecting the LEO radiation environment, including solar energetic particle (SEP) events, geomagnetically-trapped protons in the South-Atlantic Anomaly (SAA) region, and quasi-trapped and precipitating electrons from the outer radiation belt (Bruno et al., 2022; Blum et al., 2024; Vidal-Luengo et al., 2024a, 2024b; Freund et al., 2024, 2025).

CALET contains three primary detectors: a charge detector (CHD), an imaging calorimeter (IMC), and a total absorption calorimeter (TASC). The sensors used in this analysis are the topmost two layers, CHDX and CHDY, and the bottommost layer of the IMC, IMC4. The thresholds of the CHDX, CHDY, and IMC4X respectively are >1.5 MeV, >3.4 MeV, and >8.2 MeV for electrons, and >17 MeV, >37 MeV, and >52 MeV for protons. Other detector layers are excluded from these observations due to their thresholds changing depending on the ISS orbit. It is important to note that the CALET count rates cannot distinguish between electrons and protons. However, using contextual interplanetary and magnetospheric observations from other spacecraft and accounting for the ISS geomagnetic location, we are able to conclude whether certain observations are proton or electron dominated. Further details on CALET can be found in the supporting information.

## 3 Observations

### 3.1 Solar and Geomagnetic Activity

Figure 1 shows the temporal variation of the CALET count rates, along with interplanetary and geomagnetic data, between May 1 and October 1, 2024. Panel a) is the 5-min resolution integral proton intensities measured by the *Geostationary Operational Environmental Satellite-18* (GOES-18) at >10 MeV, >50 MeV, >100 MeV and >500 MeV. Panels b), c) and d) display the 6-hour resolution count rates (color code) recorded by the CHDX, CHDY and IMC4X detectors as a function of L. Finally, panel e) shows the IMF magnitude (black) and the solar-wind speed (red) based on 1-min resolution OMNIWeb (Papitashvili & King, 2020) data, and panel f) displays the IMF Bz component in GSM coordinates (black) along with the Dst index (red).

The sudden storm commencement on May 10 at 17:05UT was marked by a substantial increase in GOES >10MeV protons (not detectable by CALET). This was associated with the arrival of the merged ICME structure discussed in Section 1, and was marked by a sharp increase in the solar-wind speed, from ~450 to ~700 km/s, with Dst reaching a minimum of -412 nT. Simultaneously, the IMF exhibited a quick, highly-variable southward turning, resulting in a persistent large-amplitude southward Bz component which was most likely responsible for the superstorm and its complex time evolution. The extreme solar-wind dynamic pressure caused a strong compression and erosion of the geomagnetic field, pushing the magnetopause inside the geosynchronous orbit (Tulasi Ram et al., 2024).

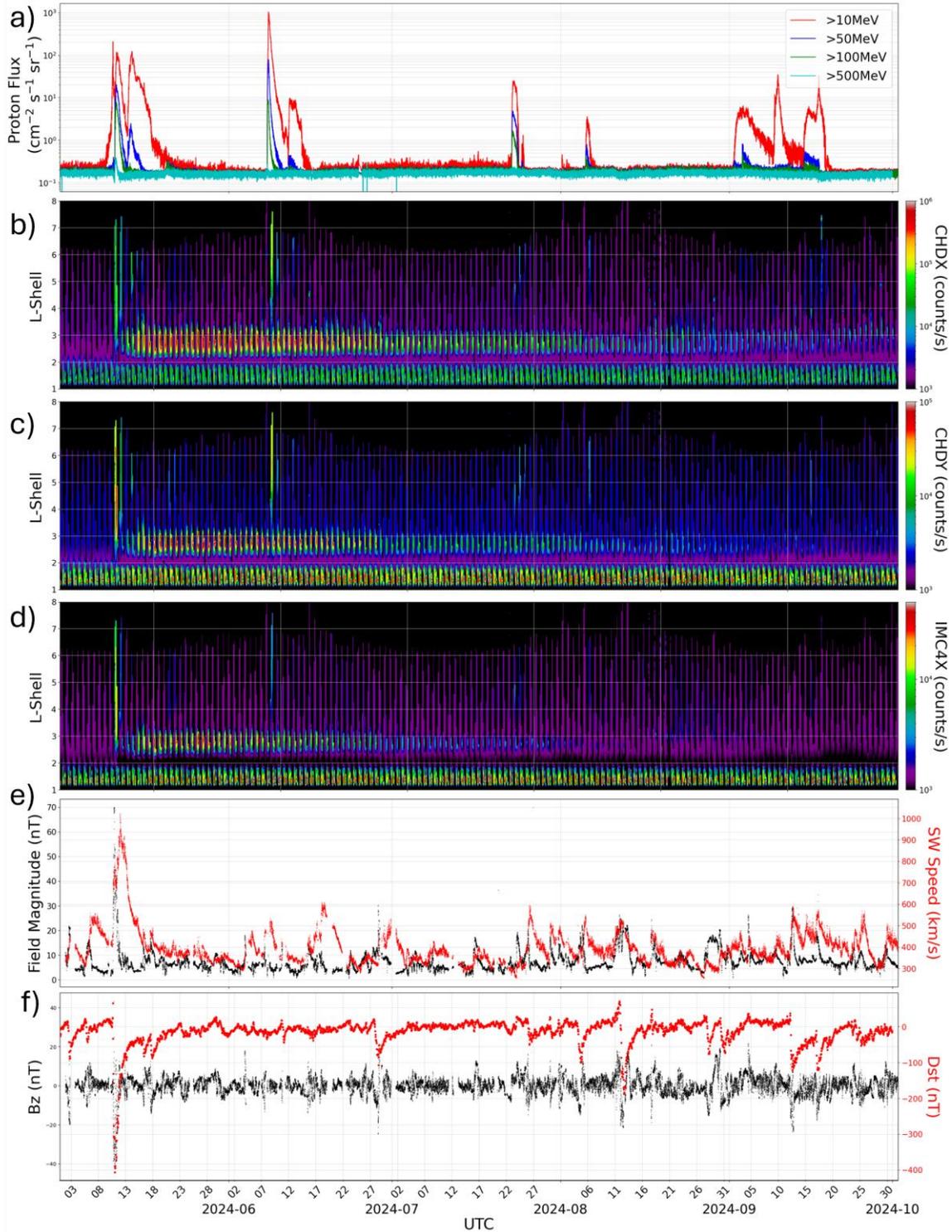

*Figure 1*: (a)Time profiles of the GOES integral proton flux for >10 MeV, >50 MeV, >100 MeV, and 500 MeV; count rates vs. L measured by (b) CHDX (>1.5 MeV electrons and >17 MeV protons) , (c) CHDY (>3.4 MeV electrons and 37 MeV protons) , (d) IMC4X (>8.2 MeV electrons and >52 MeV protons); (e) IMF magnitude and solar-wind speed; (f) IMF Bz component and Dst index.

A high-energy SEP event occurred on May 11 associated with an X5.8 flare and 1263 km/s- CME first observed at 01:36 UT. Detected by both space- and ground-based detectors, it was classified as the 74th ground-level enhancement (GLE 74), the second of solar cycle 25. The proton signal associated with this SEP event is clearly visible in the GOES >500 MeV channel and all CALET count rates, exhibiting a significant increase at moderate-to-high magnetic latitudes (L $\gtrsim$ 2.5). The enhancement can still be seen in the CHDX and CHDY data on the 12th of May, although at a narrower range of L-shell, but is no longer evident in the IMC4X, consistent with the softening of the SEP-event spectrum with time. Another small SEP event was detected on May 13 at ~11 UT, linked to the 1456 km/s CME first observed at 09:12 UT, characterized by a relatively-small increase in the >100 MeV proton intensities measured by GOES, and in the CHDX and CHDY count-rates above L~4.5.

On June 8$^{th}$, another relatively-high-energy SEP event was detected, associated with the M9.7-class flare and the 1106 km/s CME first observed at 01:53 UT. This SEP event can be seen in both the CHDX and CHDY, with a faint signal in the IMC4X. A severe (G4) geomagnetic storm occurred on June 28 at ~10 UT, due to the prolonged period of southward Bz following the arrival of an ICME erupted on June 25, with the Dst index reaching a minimum of -107 nT. This event caused an abrupt drop of the outer-radiation-belt intensities, followed by a period of relative stability.

On July 23 CALET observed an SEP event associated with a major eruption occurring well behind the western limb. The late-July period was also characterized by a gradual decrease in the measured count rates in all CALET layers, until the relatively-strong storm ($Dst_{min}$=-100 nT) occurring on August 4 at 12 UT. The storm was due to the passage of an ejecta structure probably linked to two CMEs with similar speed and direction erupted from AR 13768 on late July 31 and early August 1.

Subsequently, a new G4 geomagnetic storm occurred on August 12, with the Dst index reaching a minimum of -188 nT at 16:30UT. Later, two moderate storms were recorded on August 17 ($Dst_{min}$=-76 nT) and August 31 ($Dst_{min}$=-73 nT), triggered by the arrival of the CME most-likely ejected on August 23, and of a high-speed stream from positive-polarity coronal holes. This interval was also characterized by an increase of the CHD rates around L~3.

Finally, GOES detected some minor SEP events on September 1, 3, 9 and 14, barely visible in the CALET data. Disturbed geomagnetic conditions were recorded between September 12-13 following the arrival of an ICME likely erupted on September 10, with the Dst index reaching -121 nT. This event caused a drop in CHD rates around L~3, that had been gradually increasing in the first half of the month. A similar storm occurred on September 17 due to another ICME arrival at Earth. CALET CHD rates show a peak at L>4.5 associated with passage of an interplanetary shock, followed by a new gradual increase of the radiation-belt intensities around L~3.

### 3.2 Relativistic-Electron Observations

A remarkable feature in the CALET observations is the appearance, just after the storm main phase, of a long-lasting enhancement in count rates for all detectors in a narrow band of L-shells between 2.2 and 3.2. Given the relatively-high detection thresholds compared to typical proton energies in this magnetospheric region, we conclude that this new outer-radiation-belt component is associated with (ultra)relativistic electrons, consistent with observations by REPTile-2 and EPT during this period (Li et al., 2025; Pierrard et al., 2024). This population was found to extend well-below the so-called impenetrable barrier of L=2.8 (Baker et al., 2014). Similar to other major storms observed in the past (e.g., Baker et al., 2013, 2019), the formation of a highly-energetic electron "storage ring" can be ascribed to the extreme erosion of the plasmasphere, allowing for local acceleration of electrons at relatively-low L-shells over short timescales due to wave-particle interactions (e.g., Shprits et al., 2006; Thorne et al., 2013), followed by subsequent outward expansion of the plasmasphere beyond this storage ring location. Plasmaspheric hiss can then act to efficiently scatter relatively-low-energy ($\lesssim$1 MeV) electrons into the atmosphere, while it's less efficient at scattering higher-energy electrons (Ripoll, 2015; Reeves, 2016), resulting in an MeV and multi-MeV electron remnant belt (Li et al., 2025). A similar trend has not been observed since an analogous compression of the outer belt occurred following the Halloween storm of 2003, when the slot region was filled with relativistic electrons for about a month based on the 2-6 MeV measurements made by SAMPEX (Baker et al., 2004). However, as discussed in Section 4, the new radiation-belt feature created during the

2024 May event persisted for a much longer time.

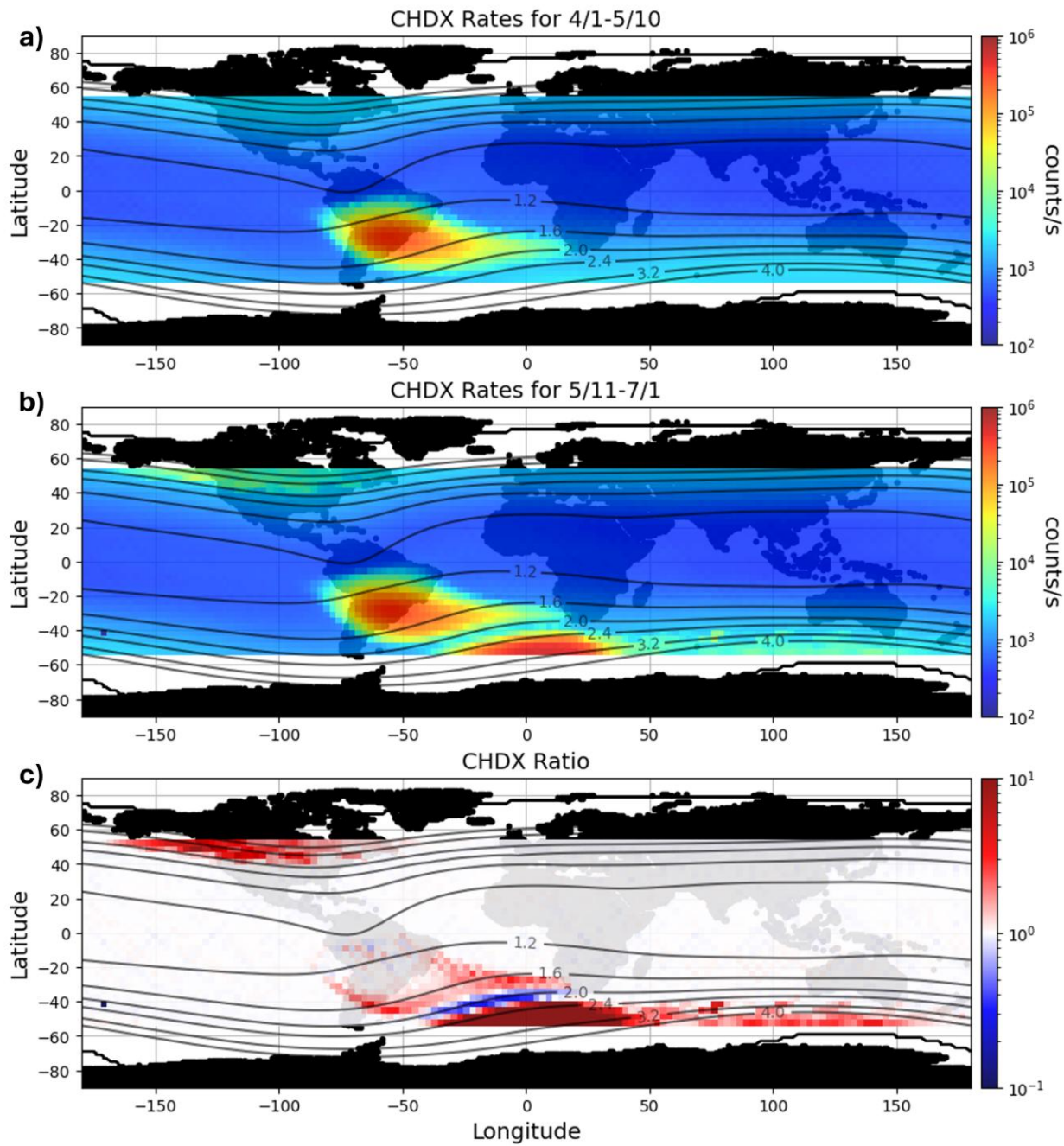

*Figure 2: (a) Count rates of the CHDX (>1.5 MeV electrons, >17 MeV protons) as a function of geographic longitude and latitude for 4/1 through 5/10. (b) CHDX count rates as a function of geographic longitude and latitude for 5/11 through 6/30. (c) Ratio of the count rates of the CHDX before and after the storm.*

The resulting radiation belt configuration is evident in Figures 2a and 2b, comparing the geographic distributions of the CHDX (>1.5 MeV) count rates during ~one-month long periods before and after the superstorm. Note that the latitudinal extent is constrained by the ISS orbit inclination (51.6°). In addition to the two moderate high-latitude enhancements in both northern and southern hemispheres mostly due to SEPs, a new structure is visible southeast of the SAA (around 0° longitude), associated with the relativistic electron population formed between L=2.2-3.2. As demonstrated in Figure 2c, displaying the ratio between the CHDX (>1.5 MeV) count rates shown in Figures 2b and 2a, the corresponding count rates increased by ~one order of magnitude.

### 3.3 SAA-Proton Observations

Another interesting aspect in Figure 2 is the change in shape and intensity of the SAA, dominated by stably-trapped protons, which sharpens along its eastern edge compared to pre-storm times. The comparison with REPTile-2 on CIRBE and EPT aboard PROBA-V measurements suggests that these effects may be due to the trapping of protons injected during the May 10-11 SEP events (Pierrard et al., 2024; Li et al., 2025; Zhang et al., 2025). In fact, the strongly-compressed magnetospheric conditions during the storm resulted in a significantly-weaker geomagnetic shielding, allowing solar protons to penetrate deeper into the magnetosphere (e.g., Selesnick et al., 2010). The changes due to this injection of protons are seen more clearly in Figure 2c, where the count-rate increase across the southeastern edge of the SAA below L=2 closely matches the proton enhancements measured by EPT. Another notable feature is the count-rate deficit in the slot region between the SAA and outer belt (L~2.0-2.2), shown by the blue structure. While Figure 2 is based on the CHDX data (>17 MeV protons), a similar trend is also observed for the CHDY (>37 MeV protons) and the IMCX4 (>52 MeV protons) count rates, with relative differences decreasing with increasing energy. As discussed by Xu et al. (2025), this depletion was caused by rapid proton losses (e.g., magnetic curvature scattering) at the outer boundary of the SAA during the main phase of the storm (e.g., Selesnick et al., 2010; Engel et al., 2016), and it was followed by a slow recovery of proton intensity levels extending to several months.

### 4 Temporal Evolution of the New Radiation-Belt Electron Component

Figure 3a shows CHDX (>1.5 MeV), CHDY (>3.4 MeV), and IMC4X (>8.2 MeV) count rates smoothed using a four-day rolling average (three-hour steps) , across the region of L=2.8 to L=3.2, where the enhancement was the highest. Following the May storm, an increase is seen in all layers, peaking~10 days later. A month of steady decay follows, before a significant depletion occurs beginning on June 28, concurrent with the next large geomagnetic storm (Dst$_{min}$ ≲ -100 nT). Because of the rolling average, the apparent dropout precedes the storm onset by ~3 days.. The remainder of July shows a smooth decay, albeit at a slower rate than observed in May/June. Finally, in August we observe two smaller dropouts associated with the two geomagnetic storms on the 4$^{th}$ and the 12$^{th}$, followed by an increase in the electron population before subsequent decay through the end of September. By the end of September, two strong storms on September 12 and 17, IMC4Xreturn CHDY and IMC4X to pre-storm background levels, while the CHDX, remains enhanced.

Figure 3b-3dd again show the smoothed averages of the CHDX, CHDY and IMC4X, but now for bins of L-shell from 2.2 to L=3 to investigate the temporal evolution of the electron population in the slot region. The largest count rates were typically observed at higher L-shells, with each bin above L=2.6 peaking at similar values. For comparison, panel 3e displays the Dst profile. At the time of the late-June geomagnetic storm, the difference in the L-shell ranges becomes apparent, with the higher L-shell ranges exhibiting a significantly larger drop in count rates when compared to the lower L bins. This trend continues for the storms in August. Similarly, when an enhancement of particles occurs in the second half of August, the higher L-shell ranges exhibit larger increases in count rates. The four L-shell regions below L=2.6 react to storms

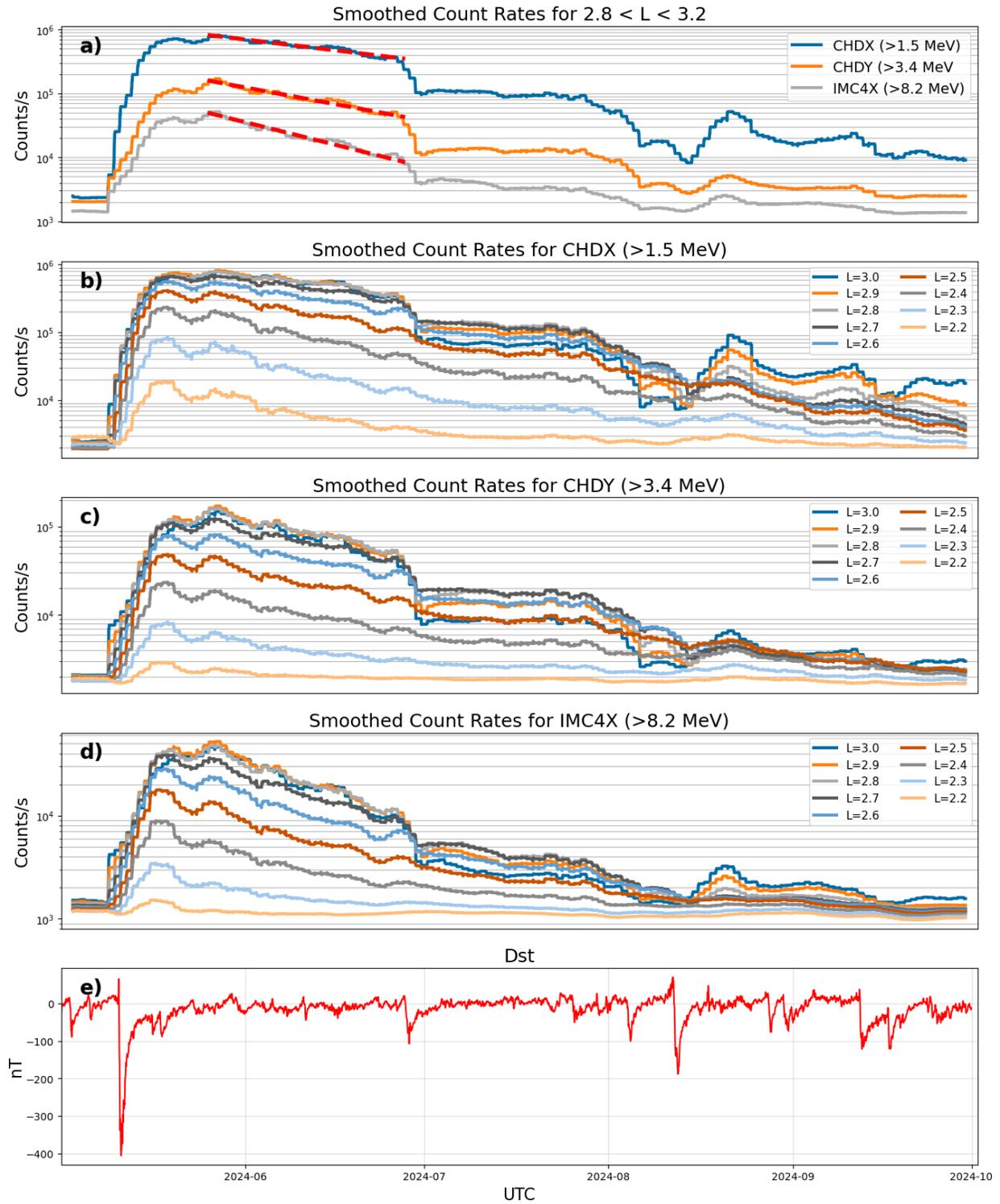

Figure 3: (a) Smoothed count rates measured by CHDX (>1.5 MeV electrons), CHDY (>3.4 MeV electrons), and IMC4X (>8.2 MeV electrons) for values of L between 2.8 and 3.2. The dashed lines represent the fits for each layer. (b-d) Smoothed count rates for the CHDX, CHDY and IMC4X for nine different bins of L from 2.2 up to 3. (e) The disturbance storm index (Dst) in nT.

comparatively weakly, showing both smaller dropouts and enhancements and exhibiting a relatively-smooth decay throughout the entire period of analysis.

To determine the lifetimes of these electron populations, we fit the decay rates using an exponential, $N(t) = N_0 e^{(-t/\tau)}$ for the period of time following the full recovery of the Dst index on May 26 till the next significant disturbance associated with the June 28 geomagnetic storm. The fits are shown by the dotted lines in Figure 3a, with lifetimes of τ=41.3+/-4.2, τ=25.4+/-2.5, and τ=19.0+/-1.6 days for >1.5 MeV, >3.4 MeV, and >8.2 MeV electrons respectively.

We also performed exponential fits of the CHDX, CHDY, and IMC4X count-rate temporal profiles for each bin of L-shell shown in Figure 3b-3d, to determine electron lifetimes as a function of L and energy. Figure 4 demonstrates the results for each detector layer. The >1.5 MeV electron (CHDX) lifetimes behave differently from the other two layers, having τ values of ~38-40 days for L values >2.7, and dropping to a minimum of τ~18 days at L=2.3. In contrast, the CHDY and IMC4X reach their minimum values of τ (~25 days for CHDY and ~19 days for IMC4X) at L=3.0, while for decreased L-shell τ reaches ~45 days at L=2.3 for the CHDY and ~72 days at L=2.3 for the IMC4X.

Note that a fit was not done for the lowest bin of L-shell for the IMC4X due to lack of a statistically significant signal.

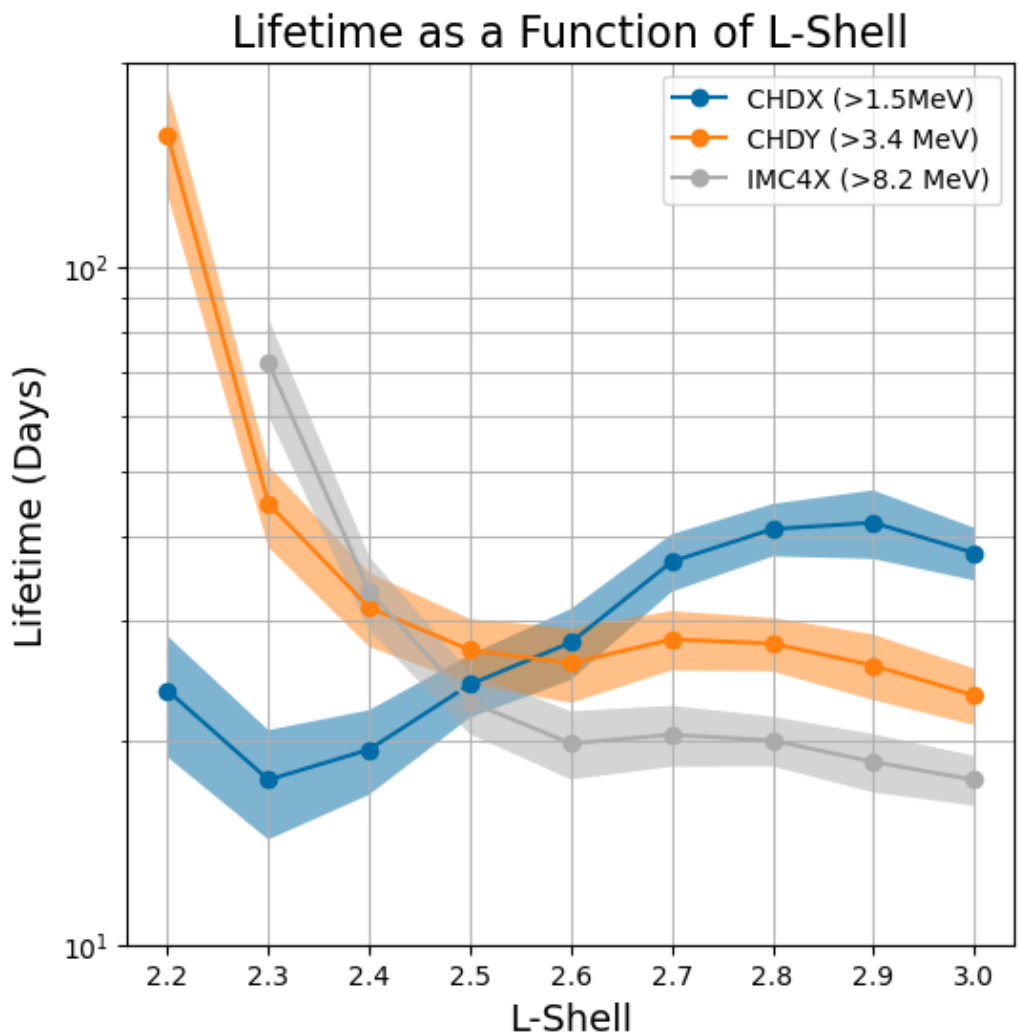

*Figure 4: Calculated electron lifetimes based on CHDX (>1.5 MeV), CHDY (>3.4 MeV), and IMC4X (>8.2 MeV) data with one-sigma uncertainty bands based on the fit-parameter errors, for bins of L from 2.2 to 3.0.*

**5 Discussion and Conclusions**

Early May 2024 was characterized by extreme space-weather activity, culminating in the May 10-11 superstorm, the most intense in two decades. Measurements by CALET enable the study of the storm's impact on the Earth's radiation belts and its evolution over several months, revealing the formation of a long-lived population of relativistic electrons injected into the slot region of L<2.8, which has not been observed since 2003. Thanks to CALET's continuous observations, we demonstrated that the enhancement of >8.2 MeV electrons lasted for well over a month, while the >1.5 MeV population persisted over 5 months following the event, significantly longer than what was observed during the 2003 Halloween storm. In addition, we calculated the effective lifetimes for the relativistic electron population between L of 2.2 and 3 to be on the order of 10s of days depending on L-shell and energy.

Electron lifetimes in this slot region are believed to be driven by losses due to wave-particle interactions as well as Coulomb collisions (e.g. Lyons et al., 1973). Hiss waves, electromagnetic ion cyclotron (EMIC) waves, VLF transmitters, and lightning-generated whistlers can all contribute to electron loss in this region (Abel and Thorne, 1998; Meredith et al., 2007, 2009; Ripoll et al., 2015). Claudepierre et al. (2019) estimated theoretical lifetimes of MeV electrons to be on the order of a few to ~100 days at L=2-3, depending on geomagnetic activity. When compared to observationally derived electron lifetimes though, they note that in general, the observations show shorter lifetimes at L<3 than those predicted by theory, suggesting potential missing physics in the models. Claudepierre et al. (2020) derived 1 MeV electron lifetimes of ~3-7 days at L=2.5-3, in agreement with observations from Benck et al. (2010), Meredith et al. (2006), and Vampola et al. (1971). Utilizing SAMPEX data, Meredith et al. (2006) found lifetimes of 2-6MeV electrons to have a minimum of $\tau$~3.5 days at L=2.5, increasing to ~20 days and 8 days at L=2 and L=3, respectively.

The overall L-shell dependence of MeV electron lifetimes calculated here is in a reasonable agreement with past observations and theoretical calculations. The energy dependence is more difficult to evaluate, particularly as observations tend to be limited in the multi-MeV range at L-shells below ~2.5-3. While Claudepierre et al. (2020) reported increasing lifetimes as energy increases at L~3, their climatology study is restricted to energies below a few MeV, so is most relevant for comparison to the CALET CHD-X channel only (>1.5 MeV electrons). They also show through theoretical lifetime estimates that scattering by EMIC waves can efficiently decrease lifetimes of electrons in the multi-MeV range around L=3, while hiss waves dominate MeV electron loss timescales near L=2. This pattern may explain the transition observed by CALET across L=2.5, where lifetimes increase with energy at lower L and decrease with energy at higher L. Another potential qualitative explanation for this contrasting trend is that collisional losses are more important at lower L, leading to longer lifetimes for higher-energy electrons, whereas wave–particle scattering and transport processes become increasingly effective at higher L, leading to shorter lifetimes for higher-energy electrons.

CALET observations provide a new opportunity to explore multi-MeV electron dynamics in the slot region following significant geomagnetic activity. The lifetimes derived using CALET are somewhat larger than those estimated by SAMPEX (Baker et al. 2004; Meredith et al. 2006). However, they are relatively-consistent with lifetimes presented in Fennel et al. (2012) who found electrons lifetimes of ~20-50 days at multi-MeV energies while observing storms during solar cycle 23 with HEO3. In addition, the CALET lifetimes exhibit a shape, as a function of L-shell, like those shown by Claudpierre et al. (2020), with the >1.5 MeV lifetimes showing a dip at L~2.3 before increasing at lower values of L. CALET results show how at higher energies the lifetime curves shift to be strictly increasing with decreasing L-shell. One important consideration for these CALET lifetime estimates is that they are derived from integral energy channels. Claudepierre et al. (2020) demonstrated that wide or integral energy channels are often affected by combined effects of highly-energy-dependent electron decay rates. Specifically, initial decay stages may be dominated by lower energies, while later decay rates show more influence from higher energies, consistent with two-stage decay rates observed by Fennel et al. (2012).

This study demonstrates the importance of the space-weather capabilities of CALET on the ISS, whose continuous monitoring of the LEO radiation environment uniquely enables investigation of short- and long-term variations in the inner magnetosphere.

**Data Availability Statement**

The CALET data used in this analysis was provided by the Waseda CALET Operation Center at Waseda University, and is publicly available via DARTS at ISAS/JAXA (https://data.darts.isas.jaxa.jp/pub/calet/). GOES proton data were obtained from the Integrated Space Weather Analysis (ISWA; https://iswa.gsfc.nasa.gov/IswaSystemWebApp/) system of the NASA Community Coordinated Modeling Center (CCMC). The DONKI system is available at https://kauai.ccmc.gsfc.nasa.gov/DONKI/. The Lockheed Martin's SolarSoft system is available at http://www.lmsal.com/solarsoft/. Interplanetary magnetic field, solar-wind speed and Dst indices data were derived from the OMNIWeb database (https://omniweb.gsfc.nasa.gov/). NOAA active region numbers are available from the NOAA Space Weather Prediction Center at https://www.swpc.noaa.gov/.

**Acknowledgements**

CALET is currently approved to continue operating through 2030. We gratefully acknowledge JAXA's contributions to the development of CALET and to the operations onboard the JEM-EF on the ISS. We also wish to express our sincere gratitude to ASI and to NASA for their support of the CALET project. The CALET space weather effort in the United States is supported by NASA/Living with a Star Science program NNH20ZDA001N-LWS and NASA's H-SR award #80NSSC21K1682 as well as by NASA Grants NNX16AC02G, NNX16AB99G, NNX11AE06G, 80NSSC20K0397 and 80NSSC24K1883 at Louisiana State University, and #NNH18ZDA001N-APRA18-004 at Goddard Space Flight Center. We also thank the CALET collaboration for access to the data used in this work. A.B. acknowledges support from NASA under award number 80GSFC24M0006. We thank the anonymous reviewers for their constructive comments and suggestions, which helped improve the clarity and quality of this manuscript.

**Conflict of Interest Disclosure**

The authors declare there are no conflicts of interest for this manuscript.

**References From the Supporting Information**